# SynthRAD2023 Grand Challenge dataset: generating synthetic CT for radiotherapy

Adrian Thummerer[1], Erik van der Bijl[2], Arthur Jr Galapon[1], Joost JC Verhoeff[3], Johannes A Langendijk[1], Stefan Both[1], Cornelis (Nico) AT van den Berg[3,4], Matteo Maspero[3,4]

[1] Department, of Radiation Oncology, University Medical Center Groningen, University of Groningen, Groningen, The Netherlands;

[2] Department of Radiation Oncology, Radboud University Medical Center, Nijmegen, The Netherlands;

[3] Department of Radiotherapy, University Medical Center Utrecht, Utrecht, The Netherlands;

[4] Computational Imaging Group for MR Diagnostics & Therapy, University Medical Center Utrecht, Utrecht, The Netherlands;

## Abstract

### Purpose

Medical imaging has become increasingly important in diagnosing and treating oncological patients, particularly in radiotherapy. Recent advances in synthetic computed tomography (sCT) generation have increased interest in public challenges to provide data and evaluation metrics for comparing different approaches openly. This paper describes a dataset of brain and pelvis computed tomography (CT) images with rigidly registered CBCT and MRI images to facilitate the development and evaluation of sCT generation for radiotherapy planning.

### Acquisition and validation methods

The dataset consists of CT, CBCT, and MRI of 540 brains and 540 pelvic radiotherapy patients from three Dutch university medical centers. Subjects' ages ranged from 3 to 93 years, with a mean age of 60. Various scanner models and acquisition settings were used across patients from the three data-providing centers. Details are available in CSV files provided with the datasets.

### Data format and usage notes

The data is available on Zenodo (https://doi.org/10.5281/zenodo.7260705) under the SynthRAD2023 collection. The images for each subject are available in nifti format.

### Potential applications

This dataset will enable the evaluation and development of image synthesis algorithms for radiotherapy purposes on a realistic multi-center dataset with varying acquisition protocols. Synthetic CT generation has numerous applications in radiation therapy, including diagnosis, treatment planning, treatment monitoring, and surgical planning.

# 1 Introduction

The impact of medical imaging on oncological patients' diagnosis and therapy has grown significantly over the last decades. Especially in radiotherapy (RT), imaging plays a crucial role in the entire workflow, from treatment simulation to patient positioning and monitoring.

Traditionally, 3D computed tomography (CT) is considered the primary imaging modality in RT, providing accurate and high-resolution patient geometry and enabling direct electron density conversion needed for dose calculations and plan optimization [1]. For patient positioning and monitoring the patient before, during, and after dose delivery, 2D X-ray-based imaging has been widely adopted. 3D cone-beam computed tomography (CBCT) - often integrated with the dose delivery machine - is increasingly playing a crucial role in traditional and more advanced image-guided adaptive radiation therapy (IGART) workflows in photon and proton therapy.

A key challenge in using the clinically available CBCT is that due to the severe scatter noise and truncated projections, image reconstruction is affected by several artifacts, such as shading, streaking, and cupping. As a result, CBCT is insufficient to perform accurate dose calculations or replanning. Consequently, patients must be referred to a repeated CT when significant anatomical differences are noted between daily images and the planning CT [2]. As an alternative, image synthesis has been proposed to improve the quality of CBCT to the CT level, producing the so-called “synthetic CT” (sCT) [3]. Additionally, conversions of CBCT-to-CT that enable accurate dose computations allow online adaptive CBCT-based RT workflows, improving the quality of IGART provided to the patients.

In parallel, over the last decades, magnetic resonance imaging (MRI) has also proved its added value for tumor and organs-at-risk delineation thanks to its superb soft-tissue contrast [4]. MRI can be acquired to verify patient positioning and monitor changes before, during, or after the dose delivery [5].

To benefit from the complementary advantages offered by different imaging modalities, MRI is generally registered to CT. Such a workflow requires obtaining CT and MRI, increasing the workload and exposing the patient to additional radiation, and requires registration of the images introducing additional ambiguities and uncertainties leading to increased margins. Recently, MRI-only based RT has been proposed to simplify and speed up the workflow, decreasing patients' exposure to ionizing radiation. This is particularly relevant for repeated simulations or fragile populations like pediatric patients. MRI-only RT may reduce treatment costs and workload and eliminate residual registration errors using both imaging modalities. Additionally, MRI-only techniques can benefit MRI-guided RT [6].
The main obstacle in introducing MRI-only RT is the lack of tissue attenuation information required for accurate dose calculations. Many methods have been proposed to convert MR to CT-equivalent images, yielding sCTs suitable for treatment planning and dose calculation.

Artificial intelligence algorithms such as machine learning or deep learning have become the best-performing methods for deriving sCT from MRI or CBCT. However, no public datasets or challenges have been designed to provide ground truth for this task and benchmark different approaches against each other. A recent review of deep learning-based sCT generation also advocated for public challenges to provide data and evaluation metrics for such open comparison [7].

# 2 Acquisition and validation methods

## 2.1 Overview dataset

This dataset consists of a total amount of 1080 CT and MRI/CBCT image pairs that were acquired between 2018 and 2022 in the radiation oncology departments of three Dutch university medical centers: University Medical Center Utrecht, University Medical Center Groningen, and Radboud University Medical Center. All patients in this dataset have been treated with external beam radiotherapy in the brain or pelvic region (photon or proton beam therapy). For anonymity, we will refer to the three centers with centers A, B, and C without specifying which letter belongs to which center.  This dataset is presented as part of the synthRAD challenge (synthrad2023.grand-challenge.org/), which is structured into two tasks: task 1 addresses MR-to-CT image synthesis and hence consists of MR/CT image pairs, task 2 focuses on CBCT-to-CT image translation and consists of CBCT/CT image pairs. Two anatomical regions were considered for each task: the brain and the pelvis. This dataset consists of four subsets: task 1 brain, task 1 pelvis, task 2 brain, and task 2 pelvis. Inclusion criteria were the treatment with radiotherapy and the acquisition of CT and either an MRI for treatment planning (task 1) or a CBCT for patient positioning during image-guided radiotherapy (task 2). Datasets for tasks 1 and 2 do not necessarily contain the same patients, and challenge participants can take part in each task separately. Figure 1 presents exemplary images for each task and anatomy.

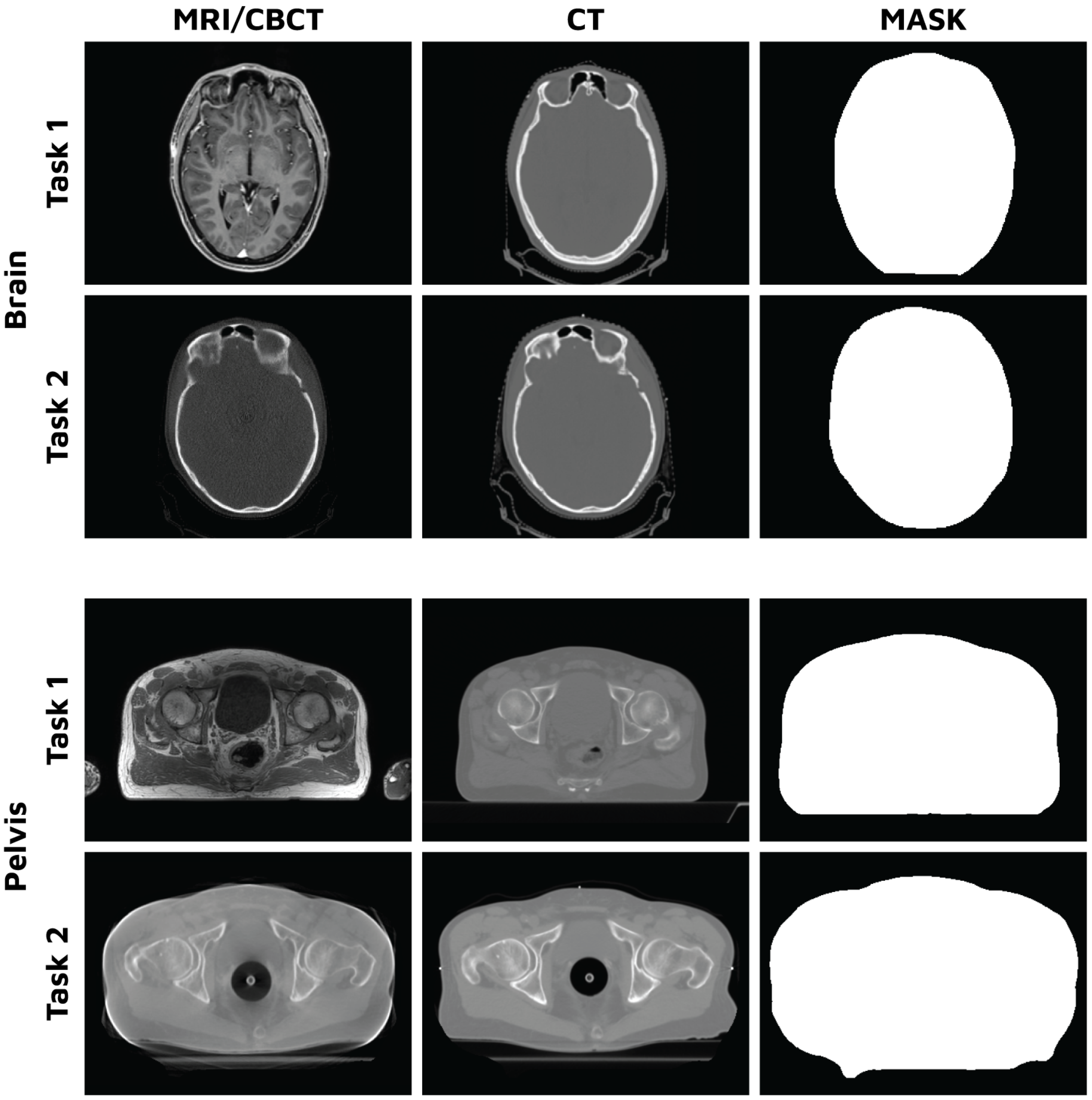

***Figure 1**: Example images for all tasks and anatomies part of the synthRAD2023 dataset. Top shows images for task 1 brain, middle-top for task 1 pelvis, middle-bottom for task 2 brain, and bottom for*

*task 2 pelvis. The first column shows the input images for the task: MRI (task 1), or CBCT (task 2); the second column is the ground truth CT, and the third column is the associated dilated body outline.*

Case selection in the brain was blind to clinical information concerning primary tumor etiology, making the tumor characteristics a random sample of the clinical routine. In the pelvis, cervical, rectal, and prostate cases were considered with an approximately equal distribution among training, validation, and test sets on an institute level. Each subset generally contains equal amounts of patients from each center, except for task 1 brain, where center B had no MR scans available. To compensate for this, center A provided twice the number of patients than in other subsets. The imaging protocols varied within and across centers. However, imaging protocols were only included if at least one-third of patients had comparable image protocols. This has been performed to preserve class balance, eliminating outliers in the contrast distribution and helping the challenge participants develop methods to handle the multi-center variability.

During data collection, no gender restrictions were considered, and the dataset consists of 64% male subjects and 36% female subjects. The shift towards more male subjects is due to the inclusion of prostate patients, making the pelvis datasets predominantly male (72.6% task 1 pelvis, 81.9% task 2 pelvis). A mostly adult patient population was collected, with patients aged 3 to 93 years and a mean age of 65. Details about age and gender distributions are presented in Figure 2.

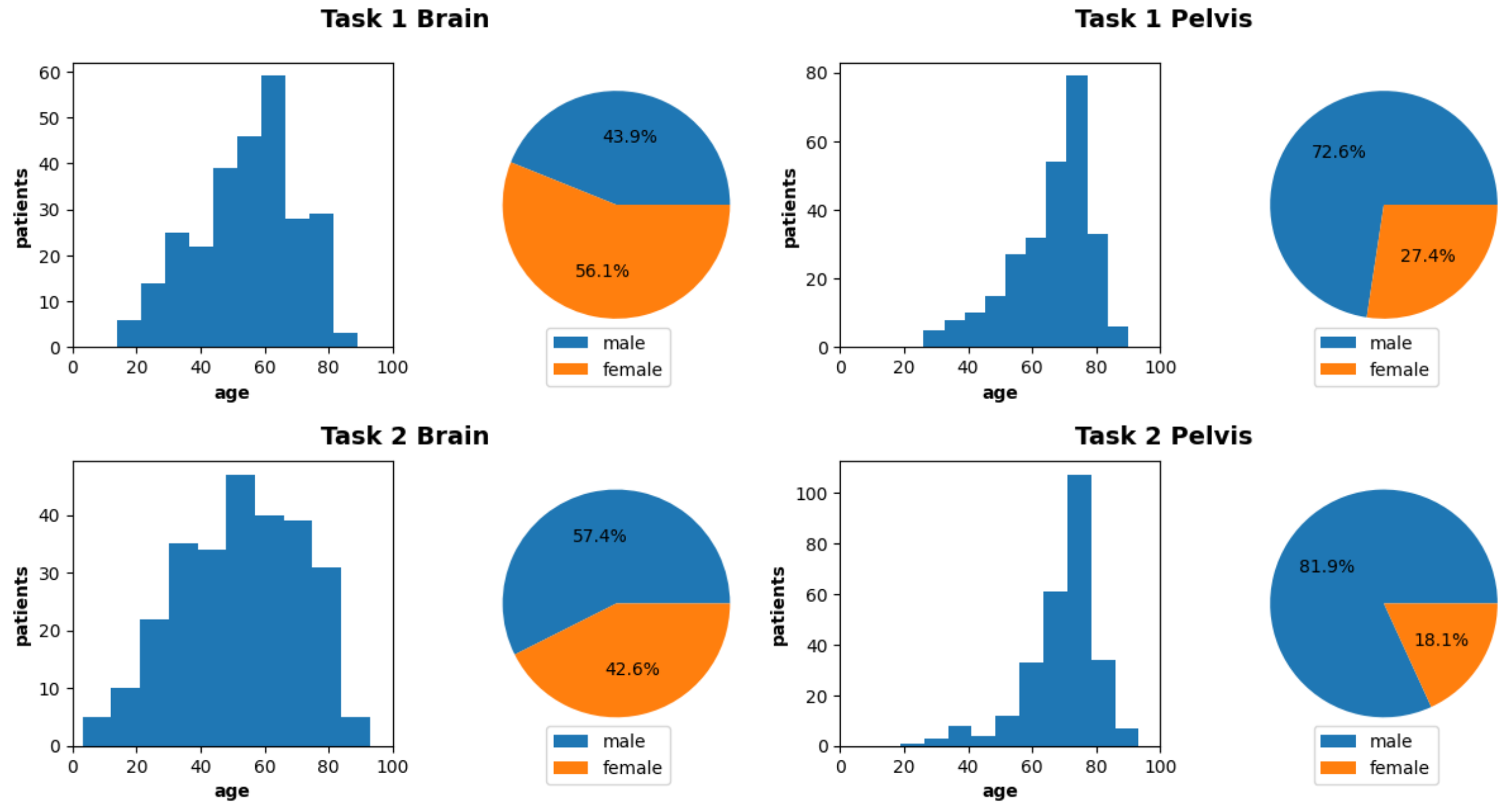

***Figure 2**: Age and gender distribution for each subset of the synthRAD2023 challenge.*

To accommodate the use of this dataset for deep learning applications and to facilitate the synthRAD2023 challenge, each subset was split into 180 training, 30 validation, and 60 test subjects as also reported in Table 1.

***Table 1**: The number of cases each institution provided per anatomy and task.*

**Train**

| | Brain | | | | Pelvis | | | |
|---|---|---|---|---|---|---|---|---|
| | Center A | Center B | Center C | Total | Center A | Center B | Center C | Total |
| **Task 1** | 60 | 60 | 60 | 180 | 120 | 0 | 60 | 180 |
| **Task 2** | 60 | 60 | 60 | 180 | 60 | 60 | 60 | 180 |

**Validation**

| | **Brain** | | | | **Pelvis** | | | |
|---|---|---|---|---|---|---|---|---|
| | **Center A** | **Center B** | **Center C** | **Total** | **Center A** | **Center B** | **Center C** | **Total** |
| **Task 1** | 10 | 10 | 10 | 30 | 20 | 0 | 10 | 30 |
| **Task 2** | 10 | 10 | 10 | 30 | 10 | 10 | 10 | 30 |

**Test**

| | **Brain** | | | | **Pelvis** | | | |
|---|---|---|---|---|---|---|---|---|
| | **Center A** | **Center B** | **Center C** | **Total** | **Center A** | **Center B** | **Center C** | **Total** |
| **Task 1** | 20 | 20 | 20 | 60 | 40 | 0 | 20 | 60 |
| **Task 2** | 20 | 20 | 20 | 60 | 20 | 20 | 20 | 60 |

Images were acquired with the clinically used imaging protocols of the respective centers for each anatomical site and reflect typical images found in clinical routine. A detailed list of acquisition details for each of the centers and subsets is provided in the following sections.

## 2.2 Task 1 (MRI-to-CT)

For task 1, MRIs were acquired with a T1-weighted gradient echo or an inversion prepared - turbo field echo (TFE) sequence and collected along with the corresponding planning CTs for all subjects.

### 2.2.1 Brain

The collected MRIs of centers B and C were acquired with a Gadolinium contrast agent, while the MRIs selected from center A were acquired without contrast.

***Table 2****: Image acquisition parameters for the* ***MRIs*** *of Task 1 Brain.*

| **Parameter** | **Center A** | **Center B** | **Center C** |
|---|---|---|---|
| Manufacturer | Philips | Siemens | Siemens |
| Model | Ingenia (89)/ Achieva dStream (1) | MAGNETOM Aera (67) /Avanto_fit (23) | MAGNETOM Avanto_fit (74) / Skyra (10) / Vida_fit (2) / Prisma_fit (4) |
| Field Strength [T] | 1.5 / 3 | 1.5 | 1.5 / 3 |
| Sequence | Spoiled T1 weighted gradient echo (turbo field echo - TFE) | Inversion prepared gradient echo (turbo field echo) | Inversion prepared gradient echo (turbo field echo) |
| Acquisition | 3D | 3D | 3D |
| Contrast | No | Gadolinium | Gadolinium |
| Flip angle [ ° ] | 8 | 8 | 8 / 9 |
| Echo numbers | 1 | 1 | 1 |
| Echo time [ms] | 3.48 - 4.06 | 2.63 - 2.67 | 1.69 - 2.97 |
| Repetition time [ms] | 7.63 - 8.67 | 1580 - 2200 | 1900 - 2200 |
| Inversion time IR [ms] | - | 900 | 900- |
| Number of averages | 1 | 1 | 1 |
| Echo train length | 224 | 1 | 1 |
| Phase encoding steps | 230 - 231 | 230 - 275 | 202 - 278 |
| Bandwidth [Hz/px] | 190 - 217 | 150 | 160 - 495 |
| Pixel spacing [mm, mm] | [0.22 - 0.96, 0.22 - 0.96] | [0.98, 0.98] | [0.98 - 1.12, 0.98 - 1.12] |
| Rows | 240 - 1024 | 236 | 224 - 256 |
| Columns | 240 - 1024 | 174 - 236 | 204 - 256 |
| Acquisition matrix | [0,232, 230-231,0] | [0,256, 230-246,0] | [0,224-256, 204-256,0] |

***Table 3****: Image acquisition parameters for the* ***CTs*** *of Task 1 Brain.*

| **Parameter** | **Center A** | **Center B** | **Center C** |
|---|---|---|---|
| Manufacturer | Philips | Siemens | Philips |
| Model | Big Bore (32) / Brilliance Big Bore (58) | SOMATOM Definition AS | Brilliance Big Bore |
| kVp | 120 | 120 | 120 |
| mA | 234 - 350 | 69 - 221 | 261 - 428 |
| Exposure | 400 - 450 | 76 - 401 | 285 - 459 |
| Exposure Time | 1143 - 1712 | 1000 | 888 - 1142 |
| CTDIvol | 42.5 - 53.5 | 6.35 - 33.3 | 33.9 - 54.5 |
| Rows | 512 | 512 | 256 - 512 |
| Columns | 512 | 512 | 232 - 512 |
| Pixel spacing [mm, mm] | [0.57-1.17, 0.57-1.17] | [0.59 - 1.27, 0.59 - 1.27] | [0.69 - 0.78, 0.69 - 0.79] |
| Slice thickness [mm] | 1 - 2 | 1 - 2 | 1 - 3 |
| Reconstruction Diameter [mm] | 294 - 600 | 302 - 650 | 350 - 400 |

### 2.2.2 Pelvis

***Table 4****: Image acquisition parameters for the* ***MRIs*** *of Task 1 Pelvis.*

| **Parameter** | **Center A** | **Center B** | **Center C** |
|---|---|---|---|
| Manufacturer | Philips | - | Siemens |
| Model | Ingenia | - | MAGNETOM Avanto_fit (n.a) /Skyra (n.a) / Vida_fit (n.a) |
| Field Strength [T] | 1.5 / 3 | - | 3 |
| Sequence | Spoiled T1 weighted gradient echo (FFE[a]) | - | Fast spin echo (T2 weighted SPACE[b]) |
| Acquisition | 3D | - | 3D |
| Contrast | No | - | No |
| Flip angle [ ° ] | 10 | - | 100 - 135 |
| Echo numbers | 2 | - | 1 |
| Echo time [ms] | 2.30 - 4.75 | - | 100 - 202 |
| Repetition time [ms] | 3.90 - 8.10 | - | 1500 - 2000 |
| Number of averages | 1 | - | 2 |
| Echo train length | - | - | 61-80 |
| Phase encoding steps | 281 - 390 | - | 197 - 262 |
| Bandwidth [Hz/px] | 400 - 1083 | - | 590 - 592 |
| Pixel spacing [mm, mm] | [0.94 - 1.14, 0.94 - 1.14] | - | [1.17 - 1.30, 1.17 - 1.30] |
| Rows | 400 - 528 | - | 288 |
| Columns | 103 - 528 | - | 384 |
| Acquisition matrix | [0,284 - 480, 284 - 480,0] | - | [384,0,0,262] |

[a]FFE= Fast field Echo; [b]SPACE = Sampling Perfection with Application optimized Contrast using different flip angle Evolution, acquired with compressed sensing;

***Table 5****: Image acquisition parameters for the* ***CTs*** *of Task 1 Pelvis.*

| **Parameter** | **Center A** | **Center B** | **Center C** |
|---|---|---|---|
| Manufacturer | Philips (178) / | - | Philips |
| | Siemens (2) | | |
| Model | Big Bore (54) / Brilliance Big Bore (124) / Biograph40 (2) | - | Brilliance Big Bore (90) |
| kVp | 120 | - | 120 |
| mA | 61 - 487 | - | 106 - 499 |
| Exposure | 51 - 599 | - | 130 - 614 |
| Exposure Time | 467 - 1332 | - | 614 - 1232 |
| CTDIvol | 3 - 35.4 | - | 7.7 - 36.4 |
| Rows | 512 | - | 512 |
| Columns | 512 | - | 512 |
| Pixel spacing [mm, mm] | [0.77-1.37, 0.77 - 1.37] | - | [0.98 - 1.17, 0.98 - 1.17] |
| Slice thickness [mm] | 1.5 - 3 | - | 2-3 |
| Reconstruction Diameter [mm] | 390 - 700 | - | 500 - 600 |

## 2.3 Task 2 (CBCT-to-CT)

For task 2, the CBCTs used for image-guided radiotherapy ensuring accurate patient position were selected for all subjects along with the corresponding planning CT.

### 2.3.1 Brain

***Table 6****: Image acquisition parameters for the* ***CBCTs*** *of Task 2 Brain.*

| **Parameter** | **Center A** | **Center B** | **Center C** |
|---|---|---|---|
| Manufacturer | Elekta | IBA | Elekta |
| Model | XVI | Proteus Plus | XVI |
| kVp | 100 - 120 | 80 | 120 |
| mA | 10 -50 | 50 | 239 - 497 |
| Exposure | - | 154 - 161 | 272 - 1176 |
| Exposure Time | 10 - 40 | 3225 | 888 - 2661 |
| Rows | 270 - 512 | 512 | 512 |
| Columns | 270 - 512 | 512 | 512 |
| Pixel spacing [mm,mm] | [0.66 - 1.17, 0.66 - 1.17] | [0.51 - 0.51] | [0.61 - 1.17, 0.61 - 1.17] |
| Slice thickness [mm] | 1 - 3 | 2.5 | 1 - 3 |
| Reconstruction Diameter [mm] | - | 260 | 310 - 600 |

***Table 7****: Image acquisition parameters for the* ***CTs*** *of Task 2 Brain.*

| **Parameter** | **Center A** | **Center B** | **Center C** |
|---|---|---|---|
| Manufacturer | Philips/<br>Siemens | Siemens | Philips |
| Model | Big Bore (56)/<br>Brilliance Big Bore (25)/<br>Gemini TF TOF 64 (2) /<br>Mx800IDT 16 (1) /<br>Biograph 40 (6) | SOMATOM<br>Definition AS | Brilliance Big Bore |
| kVp | 100 - 120 | 120 | 120 |
| mA | 20 - 358 | 69 - 158 | 10 -20 |
| Exposure | 34 - 453 | 76 - 287 | - |
| Exposure Time | 500 - 9250 | 1000 | 20 |
| CTDIvol | 0.2 - 53.5 | 6.4 - 23.8 | 22 |
| Rows | 512 | 512 | 270 |
| Columns | 512 - 800 | 512 | 270 |
| Pixel spacing [mm,mm] | [0.39 - 1.37,<br>0.39 - 1.37] | [0.58 - 1.27,<br>0.58 - 1.27] | [1, 1] |
| Slice thickness [mm] | 1 - 3 | 1 - 2 | 1 |
| Reconstruction Diameter [mm] | 203 - 700 | 302 - 650 | - |

### 2.3.2 Pelvis

***Table 8****: Image acquisition parameters for the* ***CBCTs*** *of Task 2 Pelvis*

| **Parameter** | **Center A** | **Center B** | **Center C** |
|---|---|---|---|
| Manufacturer | Elekta | Elekta | Elekta |
| Model | XVI | XVI | XVI |
| kVp | 100 - 120 | 120 | 120 |
| mA | 20 - 80 | 16 - 40 | 64 |
| Exposure Time | 10 - 40 | 25 - 40 | 40 |
| Rows | 270 - 512 | 410 | 410 |
| Columns | 270 - 512 | 410 | 410 |
| Pixel spacing [mm,mm] | [0.88 - 1.17,<br>0.88 - 1.17] | [1,1] | [1,1] |
| Slice thickness [mm] | 1 - 3 | 2 | 1 |

***Table 9**: Image acquisition parameters for the **CTs** of Task 2 Pelvis*

| **Parameter** | **Center A** | **Center B** | **Center C** |
|---|---|---|---|
| Manufacturer | Philips/<br>Siemens | Siemens/<br>GE Medical | Philips |
| Model | Big Bore (47) /<br>Brilliance Big Bore (25) /<br>Brilliance 64 (2)<br>Gemini TF TOF 64 (2) /<br>Gemini TF Big Bore (1) /<br>Biograph 20/40/64 (13) | SOMATOM Definition As (66) /<br>SOMATOM go.Open Pro (13) /<br>Optima CT580 (11) | Brilliance Big Bore |
| kVp | 100 -140 | 100 -140 | 120 |
| mA | 17 - 508 | 39 - 376 | 128 - 493 |
| Exposure | 9 - 601 | 33 - 194 | 122 - 606 |
| Exposure Time | 453 - 6162 | 500 - 1503 | 534 - 1232 |
| CTDIvol | 0.7 - 35.6 | 3.1 - 23.1 | 7.2 - 35.9 |
| Rows | 512 | 512 | 512 |
| Columns | 512 | 512 | 512 |
| Pixel spacing [mm, mm] | [0.39 - 1.37,<br>0.39 - 1.37] | [0.81 - 1.27,<br>0.81 - 1.27] | [0.98 - 1.17,<br>0.98 - 1.17] |
| Slice thickness [mm] | 0.9 - 5 | 2 - 2.5 | 2 - 3 |
| Reconstruction Diameter [mm] | 200 - 700 | 413 - 650 | 500 - 600 |

## 2.4 Preprocessing

Data preprocessing was performed to anonymize the data entirely, reduce the file size and provide the data in a more suitable file format. Preprocessing consisted of the following steps:

- File conversion
- Resampling
- Image registration
- Anonymization
- Patient outline segmentation
- Cropping

To represent the variation in a realistic multicenter setting, our preprocessing did not include any normalization or homogenization across patients or centers. All preprocessing steps were performed using python scripts in the public repository: https://github.com/SynthRAD2023/preprocessing. In the following sections, each preprocessing step is described in more detail.

### 2.4.1 File conversion

CTs, MRIs, and CBCTs were extracted as dicom files from the respective clinical databases of each institution. The dicom files were converted to a format more suitable for the synthRAD2023 challenge, namely compressed nifti (.nii.gz.). The nifti file format allows storing full 3D volumes in a single file and compressing voxel data, significantly reducing the file size.

### 2.4.2 Resampling

To have a uniform voxel grid, all images of an anatomical region were resampled to the same voxel spacing. A 1 x 1 x 1 mm3 grid was chosen for the brain, while a coarser grid of 1 x 1 x 2.5 mm3 was selected for the pelvis.

### 2.4.3 Image registration

To align the image pairs, a rigid image registration between CBCT (task 2) or MR (task 1) and resampled CT was performed using Elastix (add ref https://elastix.lumc.nl/index.php). The preprocessing repository contains Elastix parameter files for this registration. In addition, an exemplary parameter file to perform deformable registration is also provided but was not used during preprocessing.

### 2.4.4 Anonymization

By converting the images from dicom to nifti, all patient-related metadata was removed from the original files. For the brain datasets, an additional defacing of the images was required to ensure the proper anonymization of the patient. The defacing was performed utilizing the contours of the eyes and removing voxels inferior and anterior to the eyes (see Figure 3 for an example).

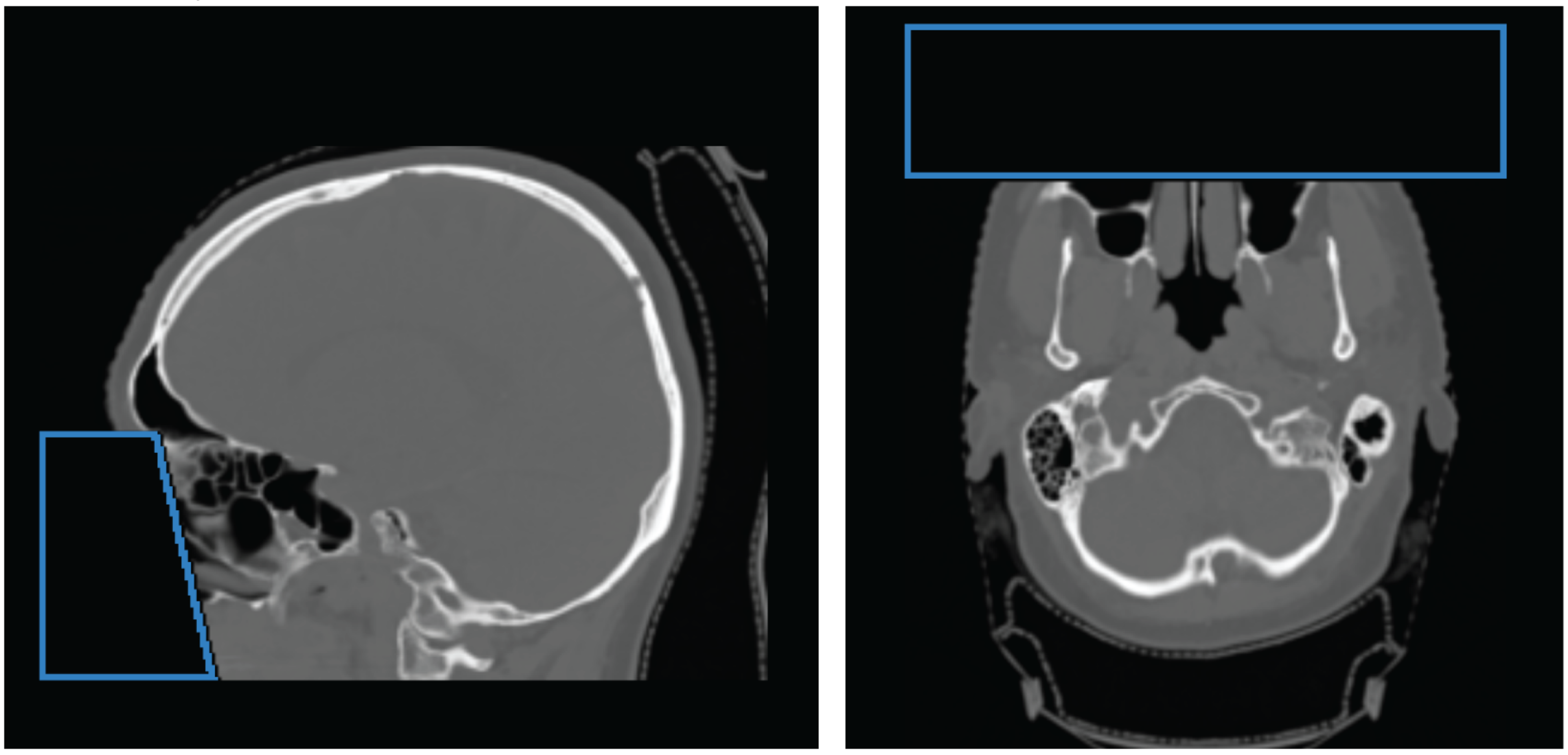

***Figure 3****: Example of a defaced brain patient. The blue ROI indicates the overwritten area with background values (-1000 for CT/CBCT, 0 for MRI) to deface the patient.*

2.4.5 Patient outline segmentation.

In addition to the MR/CBCT and CT imaging pairs, the dataset contains a binary mask of the patient outline for each case. This mask is used to ensure the same field of view on MR/CBCT and CT and is also utilized to evaluate synthetic CTs during the synthRAD challenge. The binary mask was generated using a thresholding technique and hole-filling algorithms from the ITK image processing toolkit. The resulting mask was dilated to include a margin of air surrounding the patient, which is required to calculate evaluation metrics during the synthRAD challenge.

2.4.6 Cropping

To further reduce the file size, all images were cropped to the bounding box of the patient outline, using a margin of 20 voxels.

### 2.5 Data validation

The synthRAD datasets aim to represent a realistic variation of patient characteristics and acquisition settings of the patient population. Hence, only loose inclusion criteria were necessary during patient selection, and only little validation was required. The preprocessing and data splitting (train/validation/test sets), on the other hand, required careful validation not to introduce any biases. The preprocessing results were visually checked by creating overviews containing the central axial, sagittal, and coronal slices of CBCT/MR, CT, and the patient outline mask. To assess the quality of the rigid image registration, the overview also contains images showing the difference between CBCT/MR and CT. These difference images allow a quick registration assessment but do not allow further quantification due to different intensity scales and contrasts between CBCT/MR and CT. The overview images are all included in the dataset (see dataset structure, section 3.1). Five patients showed misregistrations and required manual fine-tuning to achieve an adequate registration result.

After image registration, images were checked for abnormalities such as imaging artifacts, implants, air pockets, or variations in patient positioning. Especially in the pelvis datasets, such abnormalities were found frequently since numerous patients showed air pockets or hip implants. Significant outliers were preferably placed in the train set not to avoid having a major impact on the validation or test phase of the synthRAD2023 challenge.

## 3 Data format and usage notes

### 3.1 Data structure and file formats

An overview of the dataset structure is provided in Figure 4. On the highest level, the dataset is split into task 1 (MR) and task 2 (CBCT). Each task is then separated into the brain and pelvis anatomies. Each subset contains patient folders with a unique alphanumeric name that consists of the task number (1 or 2), the anatomy (B or P), the data providing center (A, B or C), and a three-digit patient ID. For task 1, each patient folder contains an MR (mr.nii.gz), a CT (ct.nii.gz), and a binary mask (mask.nii.gz) image. For Task 2, instead of the MR, a CBCT (cbct.nii.gz) is provided. For each anatomy, an overview folder is available containing overview images (.png), described in section 2.6, and a spreadsheet with image acquisition parameters for each patient.

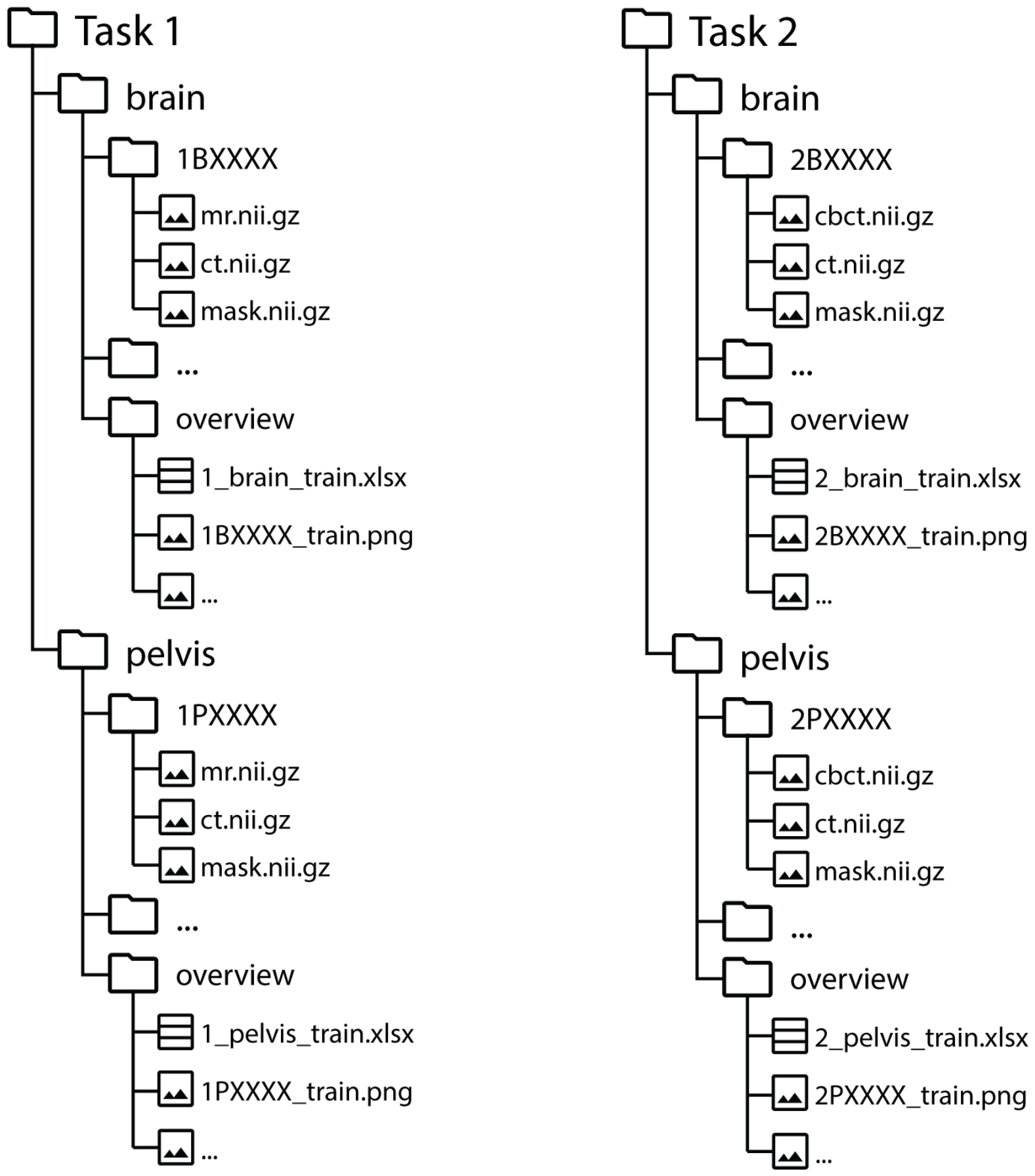

***Figure 4****: Folder structure of the synthRAD2023 dataset.*

The dataset is provided under a CC-BY-NC 4.0 International license (creativecommons.org/licenses /by-nc/4.0/) and can be downloaded from Zenodo under the following link: https://doi.org/10.5281/zenodo.7260705. The training dataset has been publicly available since April 1st, 2023. This is required for the organization of the synthRAD2023 challenge. Validation and test sets will be provided after the challenge is completed.

### 3.2 Usage notes

Compressed nifti images provided with this dataset can be read and modified using the open-source framework ITK (https://itk.org/). For various languages, e.g., Python, R, Java, and C++, a simplified interface to ITK is provided by SimpleITK (https://simpleitk.org/). Examples of how to use SimpleITK with python can be found in the preprocessing scripts. To view nifti images in a graphical user interface, 3DSlicer (https://www.slicer.org/), an open-source software for image processing, can be used.

## 4 Discussion

This dataset collection will aid in developing and evaluating synthetic CT algorithms. While numerous algorithms have been developed, the performance of these algorithms cannot be compared on a small multi-center dataset. The SynthRAD2023 dataset allows the evaluation and comparison of existing synthetic CT approaches in the pelvis and brain, and enables the development of new approaches for these anatomies.

Synthetic CT generation algorithms will benefit numerous applications such as MRI-only radiation therapy planning [6], CBCT-based adaptive radiotherapy both in an offline and online setting [ref], for patients' diagnosis [8,9,10], and surgical planning [11].

The multi-center dataset was collected to support the organization of the SynthRAD2023 Grand Challenge (https://synthrad2023.grand-challenge.org/), aiming at providing a dataset to develop rapid and automated software for patient-specific synthetic CT generation for radiotherapy purposes along with common methods for its evaluation. Specifically, we proposed to evaluate the sCT with image-based and dose-based metrics within the challenge.

The published dataset provides a heterogeneous multi-center sampling of MRI, CBCT, and CT, considering that data was acquired with independently defined positioning and immobilization guidelines using different scanners and imaging protocols. Single patient characteristics, e.g., hip implant, use of rectal balloons, tumor characteristics and presence of calcifications, also present a wide variety of conditions that may challenge sCT generation algorithms in practice. Overall, the dataset represents patients with clinical indications, providing a significant volume of patients balanced among different centers for developing algorithms that may be able to perform in clinical practice.
A limitation of the dataset is that diagnostic or other medical information is unavailable; therefore, these potentially challenging conditions are not labeled. Another limitation is that data were collected retrospectively, with reconstruction parameters limited to those used in the clinical protocol. Furthermore, raw image data was unavailable. Therefore, variations in reconstruction approaches cannot be investigated for each patient. Future dataset collections that provide raw data or high-resolution planning CT may be used to investigate the impact of noise, image reconstruction, and protocol optimization.
Time differences between CBCT/MRI and CT may lead to anatomical differences in the training and validation data, e.g., due to bladder filling, peristaltic motion, and air pockets in the rectum/bowel. Additionally, water equivalent materials, i.e., boluses, may have been positioned on the patient during irradiation even if not present during planning CT, hindering CBCT and CT correspondence.
A rigid registration was applied to overcome the misalignment between multimodality images, leaving possible deformable misalignment unresolved. After dataset inspection, we opted only to provide images aligned with rigid registration, considering that a dataset corrected for deformation is unavailable in a clinical situation where the planning CT would no longer be acquired. Considering that some sCT generation algorithms, e.g., supervised deep learning, benefit from increased data alignment, we also provided an exemplary parameter file in our pre-processing repository.

# 5 Conclusion

The SynthRAD2023 dataset will enable the evaluation and development of image synthesis algorithms for radiotherapy purposes on a realistic multi-center population, exhibiting variations in acquisition protocols. The dataset will enable a fair comparison of fully automatic approaches in medical image synthesis through the SynthRAD challenge.
Synthetic CT generation has numerous applications in radiation therapy, diagnostic tasks, and surgical planning, and the SynthRAD2023 dataset will facilitate bringing developed algorithms closer to clinical practice.

# References

1. E. S. Chernak, A. Rodriguez-Antunez, G. L. Jelden, R. S. Dhaliwal, and P. S. Lavik, The use of computed tomography for radiation therapy treatment planning, Radiology 117, 613–614 (1975).

2. S. Ramella et al., Local control and toxicity of adaptive radiotherapy using weekly CT imaging: results from the LARTIA trial in stage III NSCLC, Journal of Thoracic Oncology 12, 1122–1130 (2017).
3. Kida2018 3 S. Kida, T. Nakamoto, M. Nakano, K. Nawa, A. Haga, J. Kotoku, H. Yamashita, and K. Nakagawa, Cone beam computed tomography image quality improvement using a deep convolutional neural network, Cureus 10 (2018).
4. M. A. Schmidt and G. S. Payne, Radiotherapy planning using MRI, Physics in Medicine & Biology 60, R323 (2015).
5. J. J. Lagendijk, B. W. Raaymakers, C. A. Van den Berg, M. A. Moerland, M. E. Philippens, and M. Van Vulpen, MR guidance in radiotherapy, Physics in Medicine & Biology 59, R349 (2014).
6. M. F. Spadea∗, M. Maspero∗, P. Zaffino, and J. Seco, Deep learning based synthetic-CT generation in radiotherapy and PET: a review, Medical physics 48, 6537–6566 (2021).
7. J. M. Edmund and T. Nyholm, A review of substitute CT generation for MRI-only radiation therapy, Radiation Oncology 12, 1–15 (2017).
8. V. E. Staartjes, P. R. Seevinck, W. P. Vandertop, M. van Stralen, and M. L. Schr¨oder, Magnetic resonance imaging–based synthetic computed tomography of the lumbar spine for surgical planning: a clinical proof-of-concept, Neurosurgical focus 50, E13 (2021).
9. L. Morb´ee, M. Chen, T. Van Den Berghe, E. Schiettecatte, R. Gosselin, N. Herregods, and L. B. Jans, MRI-based synthetic CT of the hip: can it be an alternative to conventional CT in the evaluation of osseous morphology? European radiology 32, 3112–3120 (2022).
10. L. B. Jans, M. Chen, D. Elewaut, F. Van den Bosch, P. Carron, P. Jacques, R. Wittoek, J. L. Jaremko, and N. Herregods, MRI-based synthetic CT in the detection of structural lesions in patients with suspected sacroiliitis: comparison with MRI, Radiology 298, 343–349 (2021).
11. L. Morbee, M. Chen, N. Herregods, P. Pullens, and L. B. Jans, MRI-based synthetic CT of the lumbar spine: Geometric measurements for surgery planning in comparison with CT, European journal of radiology 144, 109999 (2021).